# Bayesian analysis of redistribution policy with a fixed scale

Guy Cirier*[1]. LSTA, Paris VI

**Abstract:** *A government has to finance a risk for its population. It shares the charges among the population with a fixed scale based on economic criteria. Various organisms have to collect and to redistribute fairly the subsidies. Under these conditions, when the size of the organisms is varied, the distribution's laws of the criteria are exponential families and criteria are semi linear sufficient statistics.*

## I- Generalities

Covering risks of health, fire, accidents or any other random events, is a well-known problem in insurance. For instance, price scale settled on the expenditures observed in the past is currently used in car insurance. The credibility theory [2] shows how a firm modulates the premium with a scale according to the observed claims. But in this case, each firm manages its own risks with its own scale and its own economic criteria.

Here, the problem is quite reciprocal: we suppose that a State, or some highest authority, fixes a scale for all the population and organisms managing these taxes or the repartition of the subsidies. These organisms may be districts, insurance companies or even States. We have a redistribution polity fixed by the authorities, but locally managed.

What taxpayer does not think that the charges are louder than the return, or are politically oriented? It has been seen in the past. Here, in order to avoid all discussion, we suppose a priori that the purposes and the repartition of the subsidiaries are perfectly clear and without dispute. Even in this angelic frame, what are the conditions on the share? What are the distribution's laws of the economic criteria to avoid penalizations between sub-populations?

Our approach of this economic policy is Bayesian as we have used in [3]. But here, our assumptions are more general.

## II – Definitions, hypothesis and notations

Let a population $\Theta$. Every individual of $\Theta$ is characterised by a random parameter $\theta$. The parameter $\theta$ contains all the unknown information not taken as a criterion of the risk's modulation. It has a distribution function with an a priori density $\mu(\theta)d\theta$. In the Bayesian frame used here, it is not necessary to clarify this variable or its density. We only have to suppose that it is continuously measurable. More, we are in a country where there is no tax avoidance or evasion of the population $\Theta$ when it is taxed by the fixed scale.

Now, we consider three random variables defined conditionally to $\theta$, $R|\theta$, $X|\theta$ and $Y|\theta$:

- $R$ is the risk, a random variable (r.v.) valued in $\mathbb{R}$.
- $(X,Y)$ are two random criteria. The realization $(x,y)$ modulates the expectation of $R$ to be taxed. These *r.v.* are valued in $\mathbb{R}$. Their densities are $p(x|\theta)$ and $q(y|\theta)$.

**Hypothesis H**
*The r.v. $X|\theta$ and $Y|\theta$ are independent. $p(x|\theta)$ and $q(y|\theta)$ are of class $C^2$.*

---

* guy.cirier@gmail.com

It doesn't be very coherent to suppose $R|\theta$ independent of $X|\theta$ and $Y|\theta$ if we modulate its expectation by $(x,y)$. So, its unknown density $\pi(r|\theta,x,y)$ is only supposed of class $C^2$.

**Definition and hypothesis H1 of the individual expectation**
*The individual expectation of $R|(\theta,x,y)$, $m(\theta,x,y) = E_r(R|(\theta,x,y))$, is of class $C^2$. It can be a premium or a tax covering the risk $R|(\theta,x,y)$ of an individual $\theta$ when $(X=x, Y=y)$. It is called isotropic when it is constant for all $\theta$.*

We want to modulate the expected tax of the risk $R$ with a fixed scale depending only on the realization $(x,y)$ of the criteria $(X,Y)$. But the modulation will be the same for every individual. So, it will be true for all subset $\omega \subset \Theta$.
But first, we must recall the computation of the Bayesian expectation of this tax: $m_\omega(x,y) = E_r(R|\theta \in \omega, x, y)$ with the lemma 1.
**Notations**
We note: $g(x) = p(x|\theta)$, $h(y) = q(y|\theta)$, $d\sigma = g(x)h(y)\mu(\theta)d\theta$ and $m(\theta) = m(\theta,x,y)$.

**Lemma 1**
*Under the H and H1 hypothesis, the Bayesian expectation $m_\omega(x,y)$ is written for all $\omega \subset \Theta$ :*

$$m_\omega(x,y) = \int_\omega m(\theta)d\sigma / \int_\omega d\sigma$$

As $X|\theta$ and $Y|\theta$ are independent, the density of $(\theta|x,y)$ is:
$p(\theta|x,y) = p(\theta,x,y)/p(x,y) = g(x)h(y)\mu(\theta) / E_\theta(p(x|\theta))E_\theta(q(y|\theta))$
The marginal densities $E_\theta(p(x|\theta))$ and $E_\theta(q(y|\theta))$ do not appear anymore in computation of the density of $(R|x,y,\theta \in \omega)$ because:
$p(r|x,y,\theta \in \omega)\int_{\theta \in \omega} p(\theta|x,y)d\theta = \int_{\theta \in \omega} \pi(r|\theta,x,y) p(\theta|x,y)d\theta$
So : $m_\omega(x,y)\int_\omega d\sigma = \int_\omega (\int_R r\pi(r|\theta,x,y)dr)d\sigma = \int_\omega m(\theta,x,y)d\sigma = \int_\omega m(\theta)d\sigma$
Now, we can define a "modulable" risk.

**Definitions of scale and modulable risk**
*A scale is a real not constant modulating function $f(x,y)$, independent of $\theta$, of class $C^2$.*
*The risk $R$ is modulable if any expectation $m_\omega(x,y)$ is function of $(x,y)$ only by the mean of $f$ : $m_\omega(x,y) = T_\omega(f(x,y))$ for all subset $\omega \subset \Theta$.*
**Examples**
We can take $f(x,y) = x + y$ ;
or, in the case of a single modulating statistic $x$, $f(x) = kx$ with a fixed $k$ ;
or induce recurrences $f_n(x,y) = x_n + g(x_{n-1})$

*The fact "$m_\omega(x,y)$ is modulable" induces very strong conditions on the modulating function $f$ and on the densities $p(x|\theta)$ and $q(y|\theta)$ of the modulation's criteria.*

**III- Properties of the Bayesian modulable expectation**
Let the functions: $\psi(\theta) = (f_x')^{-1}(g'/g) - (f_y')^{-1}(h'/h)$ and $\varphi(\theta) = (f_x')^{-1}m'_x - (f_y')^{-1}m'_y$. They

are of class $C^1$ except on null measure sets $\sigma$ a.e.

**Lemma 2**
*Under the previous hypothesis, if the Bayesian expectation is modulable, we have:*
$$F = \int_{\omega \times \omega} (m(\theta)[\psi(\theta_1) - \psi(\theta)] + \varphi(\theta))d\sigma(\theta) \otimes d\sigma(\theta_1) = 0$$

- We derive $m_\omega(x,y) = T_\omega(f)$ with respect to $x$, noticing that the derivative of $d\sigma$ with respect to $x$ is: $d(d\sigma)/dx = g'h(y)\mu(\theta)d\theta = (g'/g)d\sigma$. We have:
$$(dT_\omega(f)/df)f_x' \int_\omega d\sigma + T_\omega(f) \int_\omega (g'/g)d\sigma = \int_\omega ((g'/g)m(\theta) + m(\theta)'_x)d\sigma$$

- We derive $T_\omega(f)$ with respect to $y$ noticing $d(d\sigma)/dy = (h'/h)d\sigma$:
$$(dT_\omega(f)/df)f_y' \int_\omega d\sigma + T_\omega(f) \int_\omega (h'/h)d\sigma = \int_\omega ((h'/h)m(\theta) + m(\theta)'_y)d\sigma$$

- We eliminate $(dT_\omega(f)/df) \int_\omega d\sigma$ in these two equations to obtain:
$$T_\omega(f) \int_\omega \psi(\theta_1)d\sigma(\theta_1) = \int_\omega (m(\theta)\psi(\theta) + \varphi(\theta))d\sigma(\theta)$$

If we write $T_\omega(f)$ as $m_\omega(x,y) = T_\omega(f) = \int_\omega m(\theta)d\sigma(\theta) / \int_\omega d\sigma(\theta_1)$:
$$\int_\omega m(\theta)d\sigma(\theta) \int_\omega \psi(\theta_1)d\sigma(\theta_1) = \int_\omega (m(\theta)\psi(\theta) + \varphi(\theta))d\sigma(\theta) \int_\omega d\sigma(\theta_1)$$

With Fubini (see Lang p. 269), we have F. Now, we have to use the following lemma:

**Lemma 3**
*Let $f_1(\theta), f_2(\theta), g_1(\theta)$ and $g_2(\theta)$ be four real functions of $\theta$, of class $C^0$, and $w(u,v) = f_1(u)g_2(v) - f_2(v)g_1(u)$. Let $\lambda$ be a real independent of $\theta$. If we have:*
$$\int_{\omega \times \omega} w(u,v)d\sigma(u) \otimes d\sigma(v) = 0 \text{ for } \forall\, \omega \subset \Theta,$$
*Then, one of the two following conditions is verified $\sigma$ a.e.:*
- $f_1 = \lambda f_2$ and $g_1 = \lambda g_2$
- $f_1 = \lambda g_1$ and $f_2 = \lambda g_2$

$w(u,v)$ is a function of $\Theta \times \Theta$ in $\square$ of class $C^0$. If, for all $\omega \subset \Theta$: $\int_{\omega \times \omega} w(u,v)d\sigma(u)d\sigma(v) = 0$, then $w(u,v)$ is antisymmetric $\sigma \otimes \sigma$ a.e. (See Lang p. 254). But $h(u,v) = (g_2(v)g_1(u))^{-1}w(u,v)$ is also antisymmetric because:
$h(u,v) + h(v,u) = h(u,u) + h(v,v) = 0$ (As $w(u,u) = w(v,v) = 0$ by continuity of $w$). As $w$ is antisymmetric: $(g_2(v)g_1(u) - g_2(u)g_1(v))w(u,v) = 0$. Then, we have the result.

**IV- Theorem**
*Under the hypothesis H & H1, if $m_\omega(x,y)$ is modulated by the scale $f(x,y)$ and if $m(\theta)$ is not isotropic, there are two real functions $a$ and $b$ independent of $\theta$, of class $C^2$, such as:*
1- $f$ is semi linear: $f(x,y) = a(x) + b(y)$ ;
2- the densities are $p(x|\theta) = \alpha(\theta)r(x)\exp(c(\theta)a(x))$ and $q(y|\theta) = \beta(\theta)s(y)\exp(c(\theta)b(y))$
3- the expected mean must be written: $m(\theta,x,y) = c_1(\theta) + c_2(\theta)(a(x) + b(y))$

Applying lemma 3 to the formula F of the lemma 2, we obtain either $m(\theta) = \lambda$ or $\psi(\theta) = \lambda$. As $m(\theta)$ is not isotropic, $m(\theta) \neq \lambda$, then $\psi(\theta) = \lambda$ and $\varphi(\theta) = 0$.

- We derive $\psi = \lambda$ with respect to $\theta$. Because $\lambda$ and $f$ don't depend on $\theta$, we have:
$\partial \psi / \partial \theta = (f_x')^{-1} \partial(g'/g) / \partial \theta - (f_y')^{-1} \partial(h'/h) / \partial \theta = 0$
We note $c'(x) = c'_x(x,\theta) = \partial(g'/g)/\partial\theta$ and $e'(y) = e'_y(y,\theta) = \partial(h'/h)/\partial\theta$. They are of class $C^1$ $\sigma$ a.e. In $\partial\psi/\partial\theta = 0$, we substitute $(c,e)$ instead of $(x,y)$. We obtain $\partial f / \partial c = \partial f / \partial e$, a PDE well known as «transport's equation». Its general solution is $f(x,y) = t(c(x) + e(y), \theta)$.
We note: $u = c(x) + e(y)$ of class $C^2$, then, $f = t(u, \theta)$.
- As $f$ is not depending on $\theta$, then: $df/d\theta = t'_\theta(u,\theta) + t'_u(u,\theta) u'_\theta = 0$. We derive this expression, first with respect to $x$, then with respect to $y$:
$(t''_{\theta,u}(u,\theta) + t''_{u^2}(u,\theta) u'_\theta) u'_x + t'_u(u,\theta) u''_{\theta,x} = 0$
$(t''_{\theta,u}(u,\theta) + t''_{u^2}(u,\theta) u'_\theta) u'_y + t'_u(u,\theta) u''_{\theta,y} = 0$
- As $f$ is not constant, then, eliminating the factor $t''_{\theta,u}(u,\theta) + t''_{u^2}(u,\theta) u'_\theta$, we have $(u'_x)^{-1} u''_{\theta,x} = (u'_y)^{-1} u''_{\theta,y}$. But $(u'_x)^{-1} u''_{\theta,x}$ depends only on $x$ and $(u'_y)^{-1} u''_{\theta,y}$ only on $y$. So: $(c'_x)^{-1} c''_{\theta,x} = (e'_y)^{-1} - e''_{\theta,y} = d(\theta)$. By integration, we have two real functions, $a(x)$ and $b(y)$, not depending on $\theta$, such as: $c(x) = k(\theta) a(x) + k_1(\theta)$ and $e(y) = k(\theta) b(y) + k_2(\theta)$.
After calculus, as $f$ doesn't depend on $\theta$, we have: $f = a(x) + b(y)$
- Transferring this result in $\partial\psi/\partial\theta = 0$, the following equality must be constant for every $(x,y)$:
$a'^{-1} \partial(g'/g)/\partial\theta = b'^{-1} \partial(h'/h)/\partial\theta = c'(\theta)$.
Integration with respect to $x$ gives: $\partial \log(g)/\partial\theta = c'(\theta) a(x) + cte$. We have the same result for $y$. Densities are:
$p(x|\theta) = \alpha(\theta) r(x) \exp(c(\theta) a(x))$ and $q(y|\theta) = \beta(\theta) s(y) \exp(c(\theta) b(y))$.
With such densities, Jewell [5] had shown that: $m_\omega(x,y) = T_\omega(a(x) + b(y))$.
- $\varphi(\theta) = 0$ implies $(f_x')^{-1} m'_x = (f_y')^{-1} m'_y$. By substitution of variables, we have yet the transport's equation accordingly with the previous computation and leading to the result 3.
*The probability's laws must belong to the exponential family. The modulating function must be a semi linear sufficient statistic. The individual mean must be modulated by $a(x) + b(y)$.*

### V- Examination of the hypothesis H
- *If $X|\theta$ and $Y|\theta$ are not independent,* we obtain a weaker result but less interesting.
- *If the couple ($R|\theta, X|\theta$) is independent of $Y|\theta$,* the expectation of $R|(x,y,\theta)$ depends on $(\theta, x)$: $m(\theta, x)$. The computations give instead of lemma 2:
$F_1 = \int_{\omega \times \omega} \left[ m(\theta, x)(\psi(\theta_1) - \psi(\theta)) - (f_x'^{-1} m'_x(\theta, x) \right] d\sigma(\theta) \otimes d\sigma(\theta_1) = 0$
With the lemma 3, we have $\psi = \lambda$ under the condition $m'_x(\theta, x) = 0$, so $m(\theta, x) = c_1(\theta)$.
- Finally, an increasing number of criteria independent do not modify the results.

### VI - Conclusion
**Limits of economic validity**
What is the impact of these considerations in economy?
It appears natural to modulate some risks as accidents, fire, health with justified criteria, but can we extend the scope of this method to tax system or trade exchange?
In the tax system, what State expenditures can be identified to a risk?
- It is not the case for regular running costs.

- Taking a utopian State where the VAT will cover the military expenditures. Military expenditures correspond basically to a low risk, but non-negligible: the risk of war. So, the price of vegetables would be a sufficient statistic to finance these expenditures, except if we consider that the expectation of this risk is low and quite isotropic in the population.

So, the validity of the method increases with the magnitude of the risk involved for certain sub-populations.

- Lastly, when we have a few number of organisms, the continuity of functions does not remain valid. So, the situation may result from the balance of power.

**Inequalities**

- Do the experts verify if the economic criteria are sufficient statistics when they fix a scale for all a population and for all the organisms? We really don't know it.
- Organisms $i \subset I$ share a population $\Theta$ with sub-sets $\omega_i \subset \Theta$. $\omega_i$ is the portfolio of risks of the organism $i$. $\Sigma_{i \in I} \omega_i = \Theta$. If $i$ don't receive on average $m_{\omega_i}(x,y)$, the redistribution confers an advantage or a penalty for the organism. A bad distribution of the risks and of the criteria may induce serious distortions of competition $[7]$. However, regional or urban inequalities seem be accepted with resignation $[4]$.

As some well-known negative theorems in economic science $[1]$, this restrictive theorem should make us cautious with the scales fixed by highest authorities.

**References**

$[1]$ Arrow K. *Rational choice functions and ordering*. Economica. New series, Vol.26, n°102, (1959)

$[2]$ Bühlmann H. *Mathematical methods in risk theory*. NY. Springer Verlag. (1970)

$[3]$ Cirier G. *Contribution à l'étude du bonus malus*. Mémoire de l'IAF. Paris. (1987)

$[4]$ Cour des Comptes. *Situation et perspectives financières des départements*. Rapport. (2013)

$[5]$ Jewel W. *Credible means are exact Bayesian for exponential families*. Astin Bull.8. (1974)

$[6]$ Lang S. *Real analysis*. Add. Wesley, (1973)

$[7]$ Thatcher M. *Speech of Dublin*. 30 /11/1979